\documentclass[twocolumn]{aastex631}
\usepackage{breqn}
\usepackage{float}

\shorttitle{performance of a solar wind model using Synthetic magnetogram data}
\shortauthors{Henadhira Arachchige et al.}

\graphicspath{{./}{figures/}}
\begin{document}

\title{Comparing the Performance of a Solar Wind model from the Sun to 1 AU using Real and Synthetic Magnetograms}

\author[0000-0002-5592-3297]{Kalpa Henadhira Arachchige}
\affiliation{Lowell Center for Space Science and Technology\footnote{\url{https://www.uml.edu/Research/LoCSST/}}}

\affiliation{Department of Physics \& Applied Physics \footnote{\url{https://www.uml.edu/physics/}},University of Massachusetts Lowell}

\author[0000-0003-3721-0215]{Ofer Cohen}
\affiliation{Lowell Center for Space Science and Technology\footnote{\url{https://www.uml.edu/Research/LoCSST/}}}

\affiliation{Department of Physics \& Applied Physics \footnote{\url{https://www.uml.edu/physics/}},University of Massachusetts Lowell}%

\author[0000-0002-4716-0840]{Andr\'es Mu{\~n}oz Jaramillo}
\affiliation{Southwest Research Institute Boulder, Boulder, CO, USA}%

\author[0000-0002-2728-4053]{Anthony R. Yeates}
\affiliation{Department of Mathematical Sciences, Durham University, Durham, DH1 3LE, UK}%

\accepted{August 26, 2022}
\submitjournal{ApJ}

\begin{abstract}
The input of the Solar wind models plays a significant role in accurate solar wind predictions at 1 AU. This work introduces a synthetic magnetogram produced from a dynamo model as an input for Magnetohydrodynamics (MHD) simulations. We perform a quantitative study that compares the Space Weather Modeling Framework (SWMF) results for the observed and the synthetic solar magnetogram input. For each case, we compare the results for Extreme Ultra-Violet (EUV) images and extract the simulation data along the earth trajectory to compare with in-situ observations. We initialize SWMF using the real and synthetic magnetogram for a set of Carrington Rotations (CR)s within the solar cycle 23 and 24. Our results help quantify the ability of dynamo models to be used as input to solar wind models and thus, provide predictions for the solar wind at 1 AU.
\end{abstract}

\section{Introduction} 
\label{sec:intro}

The importance of space weather forecast has increased with the growth of our society's dependence on space technology. Currently, in-situ observations at 1~AU from the the Advanced Composition Explorer \cite[ ACE,][]{ACE98}, and global imaging of the Sun by, e.g., the Solar and Heliospheric Observatory \cite[SOHO,][]{SOHO95} and the Solar Dynamics Observatory \cite[SDO,][]{SDO12} provide observational constrains for a limited space weather forecast.  

To improve forecast capabilities, numerical models play a vital part in predicting solar wind conditions from the corona up to 1~AU. Early, simplified Potential Field models \citep[PFSS][]{PFSS_Altschuler_Newkirk_1969}, and the first Magnetohydrodynamic (MHD) model \citep{1971_Pneuman_Kopp} provided the first steady-state, global structure of the solar corona taking into account the solar wind and the Sun's open magnetic field that determines the Interplanetary Magnetic Field \citep[IMF,.e.g.,][]{McComas2007}. Modern, global MHD models for the ambient solar corona and solar wind extend their domain from the Sun to 1~AU, taking into account coronal heating and thermodynamics and solar wind acceleration \citep[e.g.,][]{Linker1990,Mikic1999,Usmanov2000,Odstrcil2003,cohen_2007_SC,Lionello2014,AWSoM_CoronalHeating_2014_Van_der,Merkin2016,Feng2017,Hinterreiter2019,Hazra2021}. While the modeled solar wind conditions at 1~AU have improved, and forecast can be obtain in real time, modeled forecast is still limited due to limited resolution near the Earth, and the dependence on the magnetogram input data  (most of these models require magnetogram data to constrain their inner boundary). 

The input magnetogram data heavily influences reproducing accurate, realistic solar wind predictions at 1 AU. Specifically, model predictions are only available {\it after} the magnetogram data has been acquired. In this paper, we investigate a new input for solar wind MHD models, which is derived from the 3D kinematic dynamo (Kd3) code \citep{Yeates_Andres_2013}. In Kd3, the bipolar magnetic regions (BMR) is generated by imposing velocity perturbations, where the main advantage of this code is that it can study both cycle propagation and photospheric evolution simultaneously. The code is modified to produce the surface magnetic field distributions, i.e., synthetic magnetograms. 

In this paper, we use the Threaded Field Line Model \citep[TFLM][]{TFLM_2021ApJ} and the latest version of the Alfv\'en Wave Solar Atmosphere Model \citep[AWSoM][]{AWSoM_CoronalHeating_2014_Van_der} within the Space Weather Modelling Framework \citep[SWMF][]{2012JCoPh_Toth} to predict the global coronal structure and the solar wind conditions at 1~AU using real and synthetic magnetogram data. Thus, this paper aims to demonstrate how well the synthetic magnetograms perform as input for MHD models over real magnetograms. We stress that our goal here is not to test the performance of the MHD model against real data, but rather compare its results using the real and synthetic input data. We use synthetic magnetograms produced by the Kd3 model for a number of Carrington Rotations (CR) over solar cycles 23 and 24, and test how similar the modeled conditions are comparing to case of the real magnetograms. A reasonable agreement means that the Kd3 model could provide data for future state of the Sun's phostospheric field. Thus, MHD models could provide space weather forecast for the ambient solar wind even before the magnetogram data is available.   

We describe our modeling approach and setup in Section~\ref{sec:model} and show the results in Section~\ref{sec:results}. We then discuss the usability of the synthetic magnetograms in Section~\ref{sec:discussion}, and conclude our findings in Section~\ref{sec:Conclusion}.

\section{Model Description} \label{sec:model}
\subsection{Model Description} \label{sec:Model Description}

We use the Alfv\'en Wave Solar Atmosphere Model (AWSoM) to obtain steady-state solutions in the solar corona. AWSoM serves as the Solar Corona (SC) module in the Space Weather Modelling Framework (SWMF)\citep{2012JCoPh_Toth}. Using SWMF, the SC module is coupled with a module for the Inner Heliosphere (IH), driving it's inner boundary conditions. The end result is a steady-state solution over the Carrington rotation that extends from the Sun to 1~AU. Both the SC and IH modules of are versions of the Block-Adaptive Tree Solar wind Roe-type Upwind Scheme (BATS-R-US) MHD code \citep{BATS-R-US_1999JCoPh_Powell}.

AWSoM uses Alfv\'en wave turbulence formalism to heat the solar corona and accelerate the solar wind, where the Alfv\'en wave turbulent pressure $P_A$ = ($\omega_+$+$\omega_-$)/2 is included in the momentum and energy equations, where $\omega_+$ is the energy density for the wave propagating along the magnetic field, and $\omega_-$ is the wave propagating in the opposite direction. In our simulations we use the single fluid MHD equations, even though two-temperature mode may improve the model's performance against observations. In addition to the Alfv\'en wave turbulent pressure, the model consists of detailed thermodynamics effects, such as radiative cooling and electron heat conduction, which assist in enhancing the performance of reproducing the EUV and X-ray images of the corona. The SC component uses a stretched spherical grid from 1 $R_\odot$ to 24 $R_\odot$ solar radii, and 
is using the Threaded Field Line Model \citep[TFLM,][]{TFLM_2021ApJ} to calculate thermodynamics in a one-dimensional manner very close to the inner boundary. This enables to overcome an extremely small grid size in the three-dimensional model. The IH component uses a Cartesian grid from 18 $R_\odot$ to 215 $R_\odot$, and it is driven by the 
SC solution through its inner boundary using a buffer grid between the two modules. For detailed information about SWMF and the SC-IH coupling, we refer the reader to \cite{2012JCoPh_Toth,Sachdeva2019}.

\subsection{Model inputs} \label{subsec:inputs}

AWSoM is driven by the radial magnetic field distribution on the photosphere (magnetograms). The magnetogram is used to calculate the three-dimensional potential magnetic field \citep{PFSS_Altschuler_Newkirk_1969}, which serves as the initial, non-MHD magnetic field. Alfv\'en waves energy is introduced at the coronal base in the form of pointing flux, $S_A /B_\odot$, which is a free parameter in the model. This energy heats the corona and accelerates the wind. Another free parameter in the model is the transverse correlation length of the Alfv\'en waves, $L_\perp\sqrt{B}$, which parameterizes the dissipation of the wave energy in the plasma. $L_\perp\sqrt{B}$ is responsible for the turbulent cascade caused by the partial reflection of forward propagating Alfv\'en waves \cite[see][for a complete description of the AWSoM parameters]{AWSoM_CoronalHeating_2014_Van_der}.

In this work, we optimize the values of these free parameters for each CR in order to get the best agreement with 1~AU data using observed magnetogram. We then use the same values to obtain steady-state solution driven by the associated synthetic magnetogram. Both magnetogram data are in the form of spherical harmonics, and these coefficients are calculated up to the order of 90. Table~\ref{tab:model parameter} shows the parametrization values used for each CR.\\
\\\\

\begin{deluxetable}{ccc}[htpb]
\tablecaption{Adjustable free model parameters.} \label{tab:model parameter}
\tablewidth{0pt}
\tabletypesize{\footnotesize}
\tablehead{
\colhead{} & \colhead{[$S_A /B]_\odot$ } & \colhead{$L_{\perp}\sqrt{B}$}
\\
\colhead{CR} & \colhead{($Wm^{-2}T^{-1}$)}  & \colhead{($m\sqrt{T}$)} 
}
\startdata
1925 & 6.00$\times 10^{5}$ & 3.00$\times 10^{4}$\\
1957 & 6.00$\times 10^{5}$ & 3.00$\times 10^{4}$\\
1989 & 2.60$\times 10^{5}$ & 3.00$\times 10^{4}$\\
2021 & 3.00$\times 10^{5}$ & 1.50$\times 10^{5}$\\
2069 & 6.00$\times 10^{5}$ & 3.00$\times 10^{5}$\\
2086 & 1.10$\times 10^{5}$ & 6.00$\times 10^{4}$\\
2112 & 3.00$\times 10^{5}$ & 1.50$\times 10^{5}$\\
2151 & 5.00$\times 10^{5}$ & 6.00$\times 10^{5}$\\
2164 & 3.00$\times 10^{5}$ & 1.50$\times 10^{5}$\\
\enddata
\end{deluxetable}


\subsubsection{Real magnetogram} \label{subsubsec:real}

Magnetograms are synoptic observations of the Sun's photospheric radial magnetic field. These maps provide observations for an entire disk of the sun during one solar rotation of 27.27 days, known as Carrington Rotation (CR). In this study, we use Synoptic magnetograms provided by several observatories, including the Michelson Doppler Imager (MDI), the Global Oscillation Network Group (GONG), and the Synoptic Optical Long-Term Investigation of the Sun (SOLIS) observatory. These observatories use the Zeeman effect to detect the strength and the polarity of the magnetic field at the sun. However, there are problems and errors associated with real magnetograms which might affect the final output obtained from the models. For more details about these errors, refer to section 4.1 in \cite{Magne_issue}. Table~\ref{tab:Observatory} shows the list of modeled CRs and the magnetogram data used for each CR. These CRs occurred during solar cycle 23 and 24.

\begin{deluxetable}{cc}[htpb]
\tablecaption{Real magnetogram inputs. \label{tab:Observatory}}
\tablewidth{0pt}
\tabletypesize{\footnotesize}
\tablehead{\colhead{CR} & \colhead{Observatory} 
}
\startdata
1925 & Michelson Doppler Imager (MDI) \\
1957 & Synoptic Optical Long term Investigation \\ & of the Sun (SOLIS) \\
1989 & Michelson Doppler Imager (MDI) \\
2021 & Michelson Doppler Imager (MDI) \\
2069 & Michelson Doppler Imager (MDI) \\
2086 & Global Oscillation Network Group (GONG) \\
2112 & Global Oscillation Network Group (GONG) \\
2151 & Global Oscillation Network Group (GONG) \\
2164 & Global Oscillation Network Group (GONG) \\
\enddata
\end{deluxetable}

\subsection{Synthetic magnetogram using Kd3} \label{subsubsec:synthetic}

We use the improved Kd3 code to generate surface field distribution of the sun for selected CRs to simulate the synthetic magnetogram case. Kd3 is a 3D flux transport model with a radial dimension that allows flux emergence and flux subduction to allow more realistic evolution of the magnetic field. The fundamental governing equation for Kd3 is the magnetic induction equation \ref{eqn: dynamo equation} \citep{Yeates_Andres_2013}.

\begin{equation}
\label{eqn: dynamo equation}
    \frac{\partial \textbf{B}}{\partial t}=\nabla \times(\textbf{v} \times \textbf{B}) - \nabla \times(\eta \nabla \times \textbf{B})
\end{equation}

The above equation is solved using a finite difference scheme by imposing localized velocity perturbations and prescribing turbulent diffusivity profile \textbf{$\eta$}. Active regions at the solar surface emerge out of a toroidal magnetic field at the tachocline, and the properties of these regions are reproduced by calibrating the velocity perturbations. This velocity has three components, an outward radial velocity that transports the magnetic flux from the tachocline to the solar surface, the diverging component that expands the rising tube with increasing heights, and a vortical flow to capture the net effect of helical turbulence on the rising tube. The advantage of this model is that it can generate the emergence and decaying of BMRs', which can be used to identify the sun's active regions in a CR. This method avoids the problems associated with depositing artificial flux tubes (\cite{Hazra_Nandy};\,\cite{Jaramillo_2010}) and the location of the emerging flux tubes is chosen based on the distribution of magnetic field at the bottom of the convection zone \citep{Yeates_Andres_2013}. 

\subsubsection{Bipolar Magnetic Region Database and Input in Kd3 Simulation}

We drive flux emergence in Kd3 using a Bipolar Magnetic Region (BMR) database constructed from NSO synoptic magnetograms \citep{BMRKPVT, whitbread-etal2018}.  This database contains the flux, latitude, longitude, and dipolar moment of each region as observed in each carrington rotation.

As described in \citep{Yeates_Andres_2013}, we only use the latitude, longitude, and dipolar moment to determine the moment, place, and tilt of the flux emergence.  The magnetic flux is not specified, but it is determined only by the availability of toroidal flux inside the convection zone.   At the moment, our flux emergence is completely deterministic and reflects the historical observations provided by our BMR datasets.  In future work we plan to drive flux emergence statistically based on toroidal field conditions in the convection zone.

\subsubsection{Dynamo Simulation Parameters and Setup}

We use the same simulation parameters and setup as in \cite{Yeates_Andres_2013}.  These parameters were optimized to match the spatio-temporal distribution of toroidal field at the base of the convection zone and the observed distribution of BMRs in the photosphere of subsequent cycles.   

The initial conditions for the simulations of isolated tubes are described in Section 4.1 in \cite{Yeates_Andres_2013} and are consist
of an empirical formalism of a purely toroidal field layer in the tachocline. For each isolated flux-tube, a velocity field perturbation is introduced to cause it to rise to the surface, where we match each flux-tube with the observed magnetic flux, location, and orientation of a real active region at some particular time. This way of course, the Kd3 model still relies on observations. Nevertheless, it is possible that these emergence parameters themselves could be modeled in the near future using, e.g., Machine Learning algorithms. Our work presented here aims to provide a quantitative analysis regarding the ability of such synthetic magnetogram data to predict the solar wind conditions at 1AU. 

We start our simulation at the beginning of solar cycle 20 initialized with symmetric toroidal belts with an average flux density of 250G and a weak dipolar field.  \citep[Mathematical expressions describing our initial conditions are also identical to those described in][]{Yeates_Andres_2013}.  We then proceed to run Kd3 for four cycles, seeding emergence during each cycle using the data described above.  Kd3 is not able (at the moment) to match the observed varying delays between one cycle and the next. Because of this, we shift the beginning of each cycle to match the evolution of the internal toroidal field.  However, once the cycle start has been prescribed, the time, latitude, longitude, and tilt of the BMR emergence is governed by observations.

Using the simulation parameters of \cite{Yeates_Andres_2013}, we perform a series of simulations to reach solar cycle 23.  The procedure for all of them is the same:

\begin{enumerate}
    \item Using the generic initial conditions described above, drive emergence using data from cycle 20.
    \item Let the simulation continue in order to determine the best time alignment between emergences of cycle 21 and the evolution of the future internal toroidal field.
    \item Reset the simulation, but this time run it with emergences from both cycle 20 and cycle 21 (delaying the beginning of cycle 21 emergence to the optimal time identified above).
    \item Let the simulation continue in order to determine the best time alignment between emergences of cycle 22 and the evolution of the future internal toroidal field.
    \item Repeat steps 3 and 4 for cycle 22 and 23.
\end{enumerate}

\subsubsection{Toroidal to Poloidal Field in Kd3}

The advective emergence of flux from within the convection zone by Kd3's advective “bubbles” serves two purposes: 1.\ It transports flux that is located originally within the convection zone, and 2.\ It imparts a tilt to the emerging flux bundle creating a poloidal component.  \cite{Yeates_Andres_2013} found that this poloidal field is more than sufficient to seed the toroidal field of the next cycle via meridional flow, turbulent pumping, turbulent convection, and differential rotation.  No additional source is necessary. In other words Kd3 works as a self-sustained, pure Babcock-Leighton dynamo (once we add the advective mechanism that transports flux to the surface), by shearing and subducting toroidal field generated out of the collective emergence of bipolar magnetic regions.

\subsection{Simulations} \label{subsec:simulation}

\begin{figure*}[htpb!]  
\epsscale{1.17}
\plotone{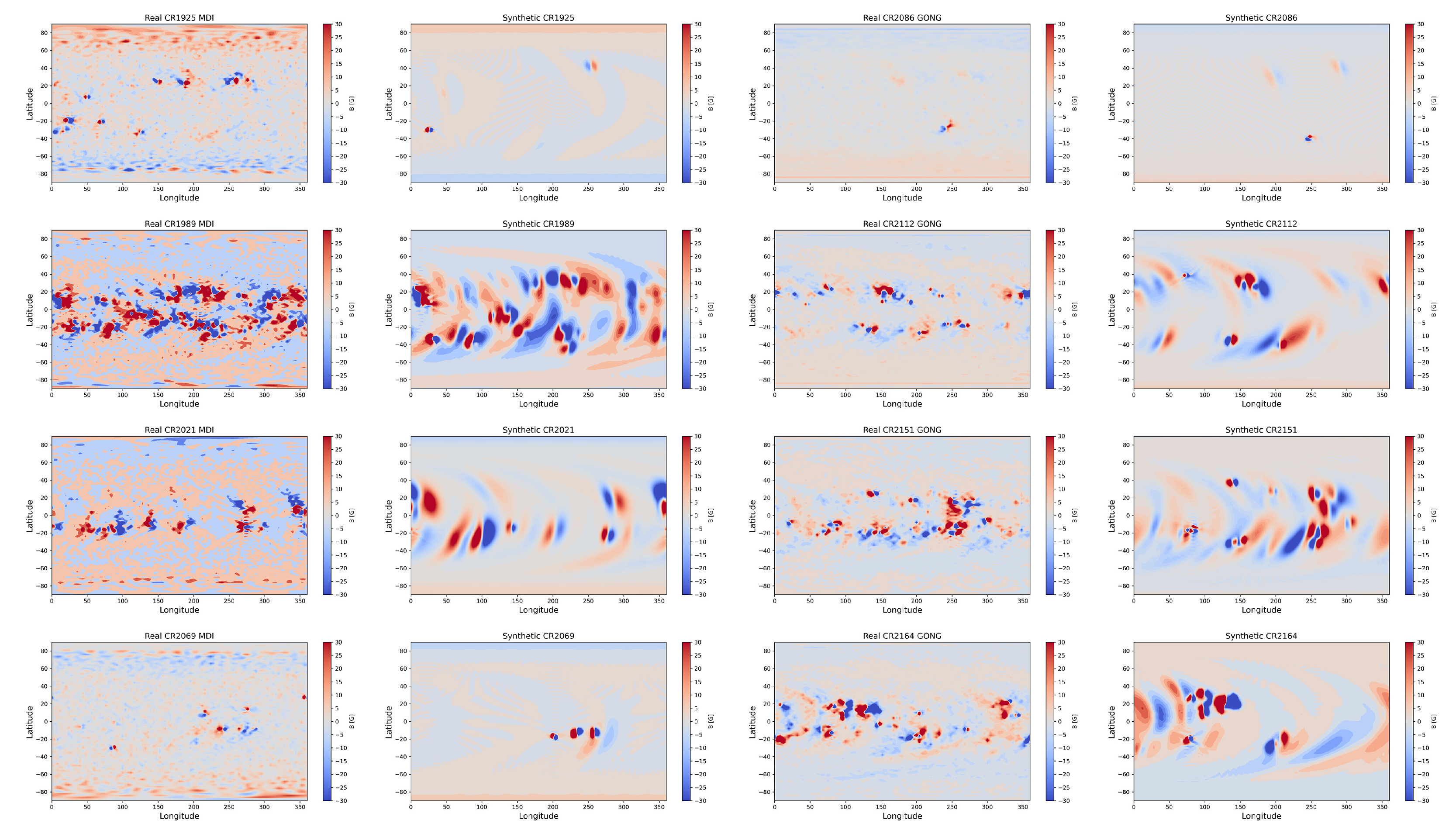}
\caption{Side-by-side comparison of the real and synthetic maps for CRs occurred in solar cycles 23 \& 24.}
\label{fig:all_mag}
\end{figure*}

For each CR, we first obtain a steady-state MHD solution using the real magnetogram input, where we modify AWSoM's free parameters to obtain the best match to 1~AU in-situ observations. We then keep the model parameters the same when obtaining a solution using the synthetic magnetogram to keep the model consistent for both cases. Side-by-side comparison of the real and synthetic maps, we refer the reader to see \ref{fig:all_mag}. For each steady-state solution from the Sun to 1~AU, the model provides two output results. First, AWSoM provides a set of synthetic EUV images for the $171$\;\AA, $195$\;\AA, and $284$\;\AA \;bands. We compare these synthetic images with the real, observed images to validate the density and temperature global structure in AWSoM. Second, we extract the solution of the IH module along the orbit of the Earth during that CR. This extraction can be directly compared against 1~AU in-situ observations of the solar wind. The output of the IH domain is the set of the MHD variables: the mass density, $\rho$, the velocity vector, ($u_x$, $u_y$, $u_z$), the magnetic field vector ($B_x$, $B_y$, $B_z$), and the pressure $p$. Using these parameters, we compare the modeled and observed solar wind number density, $n$, speed, $u$, magnetic field magnitude, $B$, and plasma temperature, $T$.


\section{Results} \label{sec:results}

We present steady-state solar wind simulations driven using real and synthetic magnetogram inputs for a selected set of CRs within the solar cycles 23 (CR 1925, 1957, and 1989) and 24 (CR 2021, 2069, 2086, 2112, 2151, and 2164). We compare the SWMF simulated results with in-situ observations at 1 AU obtained from the OMNI database\footnote{{\tt https://omniweb.gsfc.nasa.gov}} and SOHO/EIT\footnote{{\tt https://umbra.nascom.nasa.gov/eit}} images.

\begin{figure*}[htpb]  
\epsscale{1.12} 
\plotone{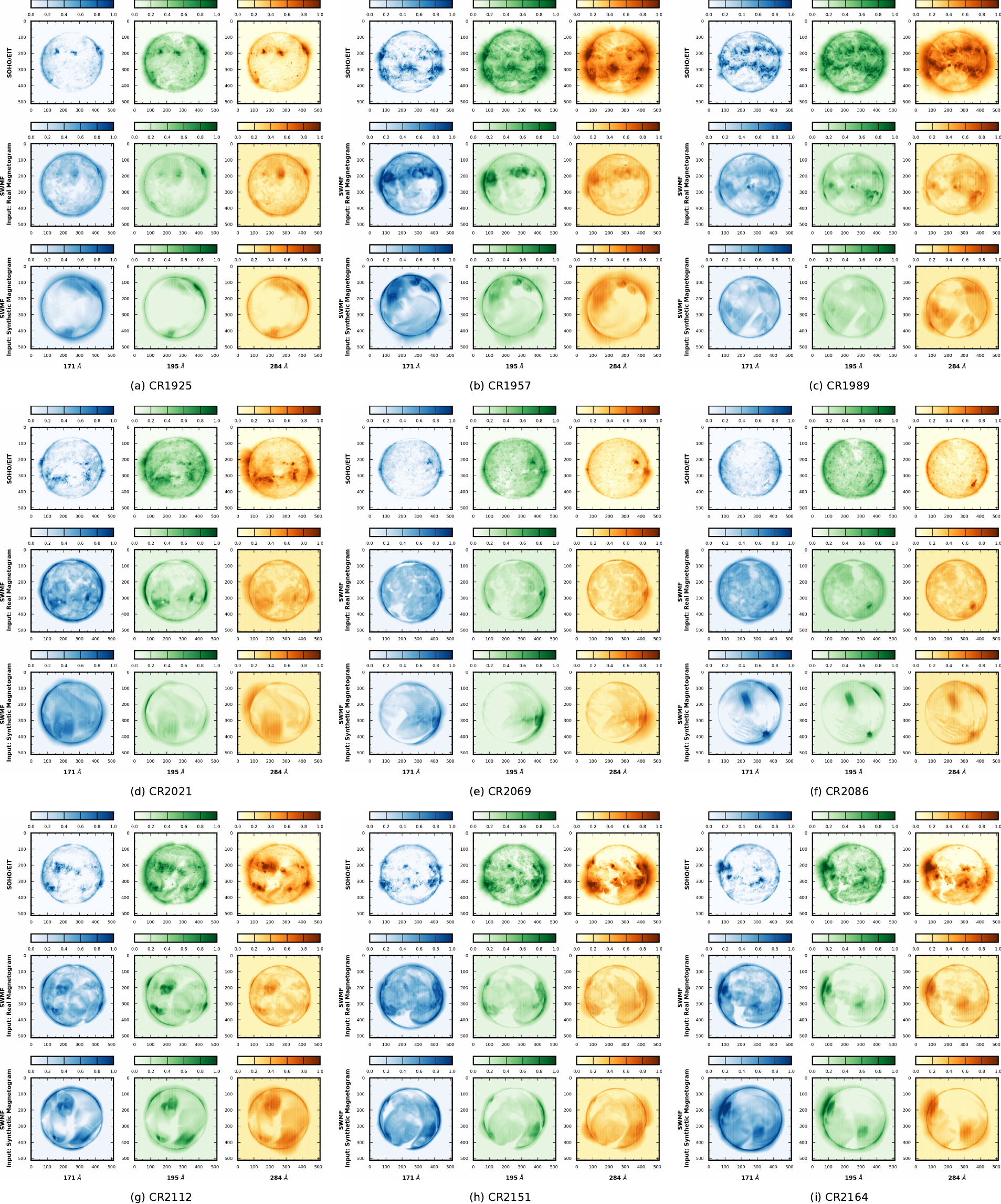}
\caption{Comparison of synthesized EUV images of the model with SOHO/EIT EUV images. The columns are from left to right for $171$\AA, $195$\AA, and $284$\AA. Top panels: observational SOHO/EIT images. Middle panels: synthesized EUV images of the model driven by real magnetogram. Bottom panels: synthesized EUV images of the model driven by synthetic magnetogram. The images  are generated for CRs, (a) 1925, (b) 1957, (c) 1989, (d) 2021, (e) 2069, (f) 2086, (g) 2112, (h) 2151, and (i) 2164.}
\label{fig:EUV_all}
\end{figure*}

\subsection{Extreme Ultra-Violet Images (EUVI)} \label{subsec:Extreme UltraViolet Images (EUVI)}

The SC steady-state MHD solution holds the steady-state density and temperature structure for the particular CR. SC has a tool to create synthetic line-of-sight (LOS) images in the EUV and X-ray bands by integrating the electron density square along the LOS, taking into account the local response function that is calculated from atomic databases \citep[e.g., CHIANTI,][]{Dare97}. For more information about the synthetic LOS images, we refer the reader to \cite{Cooper_Downs_TD_model_GSCM_2010ApJ}. For each solution, we generate synthetic LOS images as observed from the Earth during the center of the CR in the $171$\;\AA, $195$\;\AA, and $284$\;\AA\   bands. Figure \ref{fig:EUV_all}(a-i) shows the comparison for the SWMF synthesized EUV images obtained from the real and synthetic magnetogram inputs, with observed LOS images for the CRs 1925, 1957, 1989, 2021, 2069, 2086, 2112, 2151, and 2164. The observation time for all the rotations coincides with the central meridian times of the real and synthetic maps used for the simulations. Each subplot set (associated with each CR) includes the $171$\;\AA (left column), $195$\;\AA (middle column), and $284$\;\AA (right column) bands for the observed (top row), real magnetogram (middle row), and synthetic magnetogram (bottom row). Figure \ref{fig:EUV_all}(a-c) visualize the EUV image comparison for the CRs within solar cycle 23, while figure \ref{fig:EUV_all}(d-i) shows the comparison for CRs within solar cycle 24. Overall, both magnterograms do a reasonable job in reproducing the observed images, reproducing most of the large-scale bright features and coronal hole locations reasonably well. However, the synthetic manetograms produce a blurry version of the real magnetogram, as they do not capture the small-scale features that may appear in the real magnetogram. There seems to be also some small shift in the synthetic magnetograms, that may be related to the timing of the dynamo solution with the start/end time of the CR.    


\begin{figure*}[htpb]  
\epsscale{1.17} 
\plotone{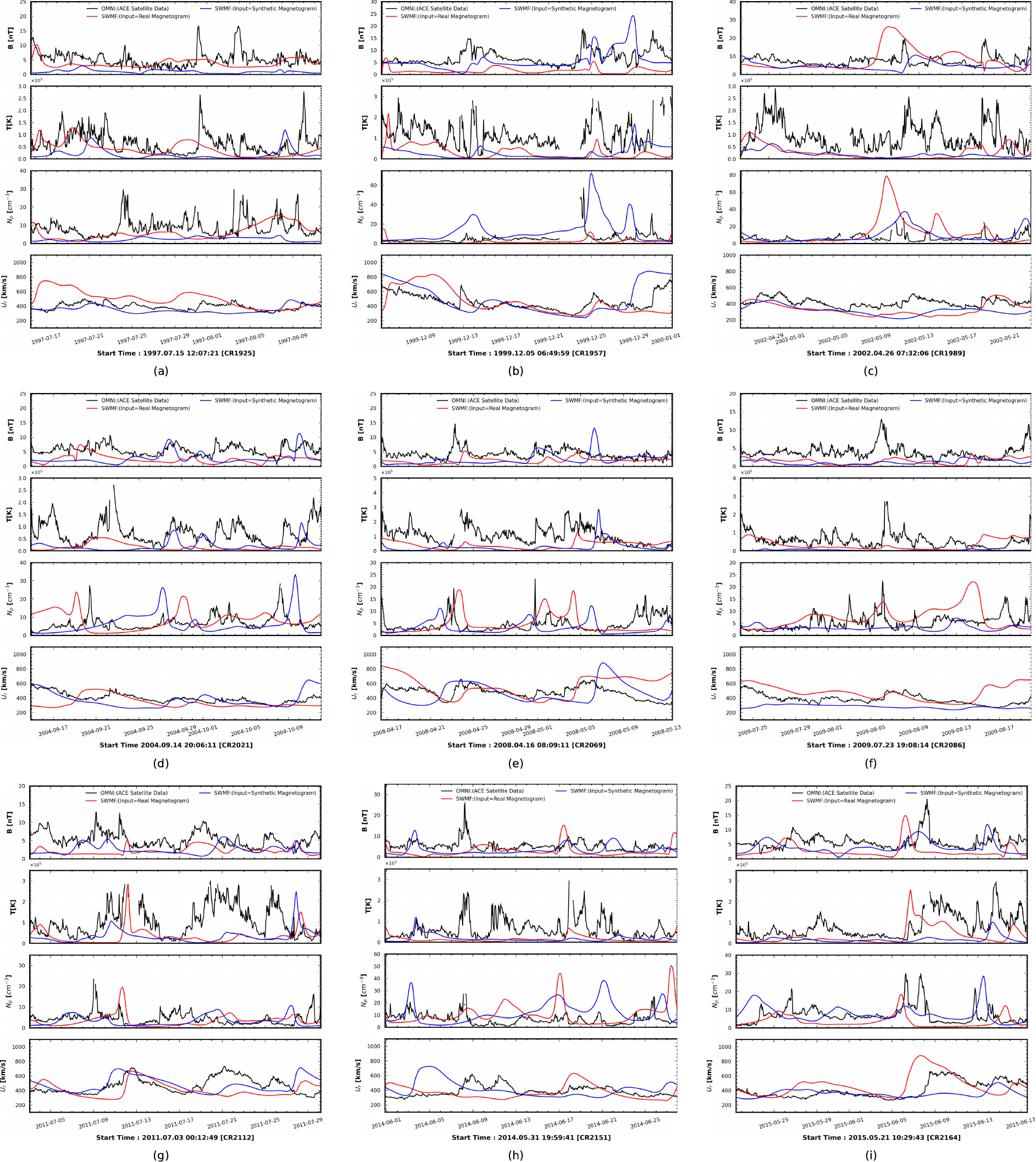}
\caption{OMNI data (black) and SWMF results for solar wind parameters driven by real magnetogram data (red) and from synthetic magnetogram data (blue) for CRs, (a) 1925, (b) 1957, (c) 1989, (d) 2021, (e) 2069, (f) 2086, (g) 2112, (h) 2151, and (i) 2164.
\label{fig:OMNI_linear}}
\end{figure*}


\begin{figure*}[htpb]  
\epsscale{1.17} 
\plotone{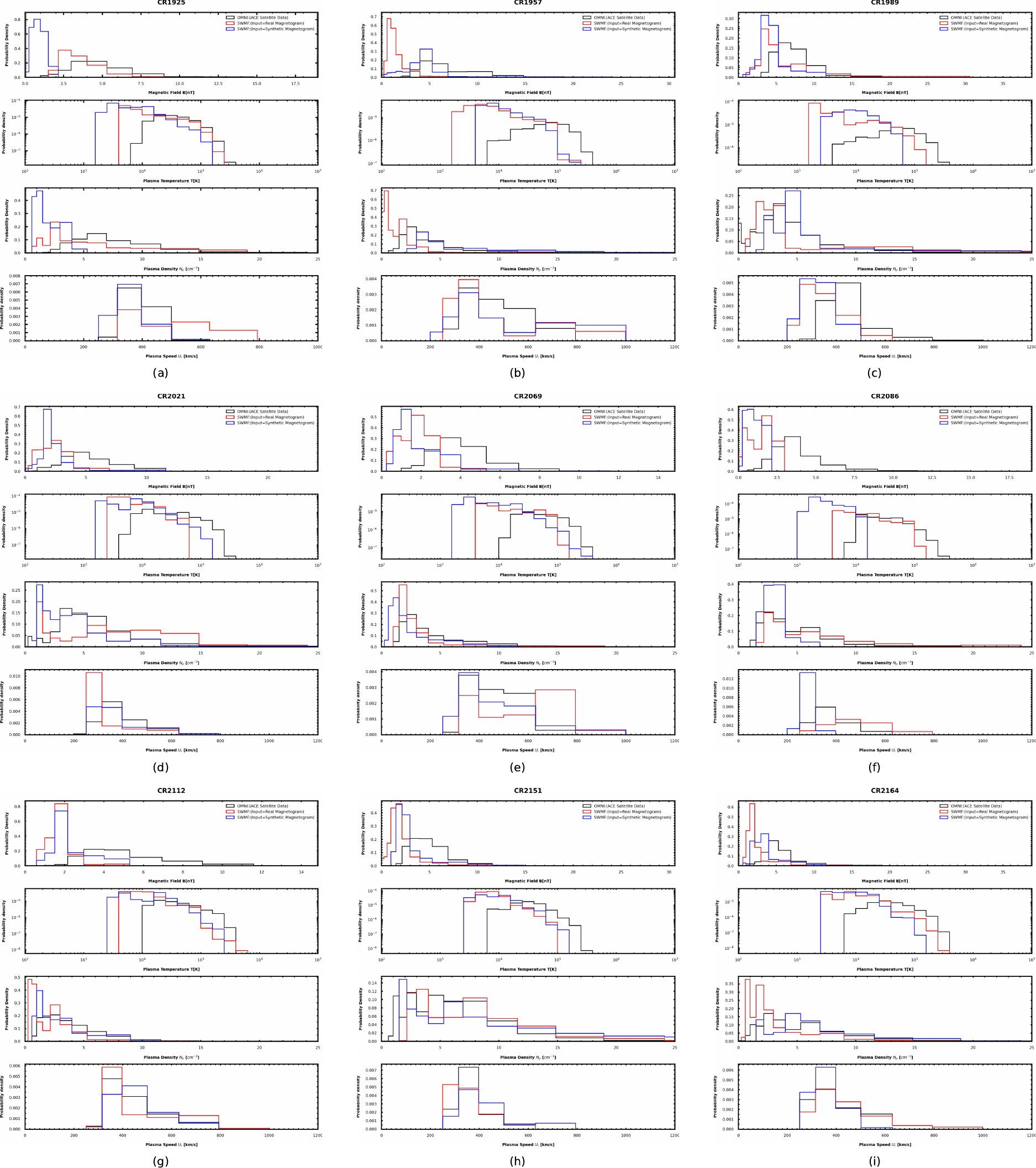}
\caption{Comparison of the distributions for solar wind at 1 AU. OMNI data (black) and SWMF results for solar wind parameters driven by real magnetogram data (red) and from synthetic magnetogram data (blue) for CRs, (a) 1925, (b) 1957, (c) 1989, (d) 2021, (e) 2069, (f) 2086, (g) 2112, (h) 2151, and (i) 2164.
\label{fig:DISTall}}
\end{figure*}

\subsection{OMNI data} \label{subsec:OMNI}

We compare the SWMF predicted solar wind parameters at 1 AU (using the two magnetogram inputs) with real in-situ observations. The in-situ observations include the hourly averaged solar wind conditions. Figure \ref{fig:OMNI_linear} (a-i) shows the comparisons of simulation results at 1 AU for the selected CRs within the Solar cycle 23 and 24. Top to bottom panels show the total magnetic field (B), plasma temperature (T), plasma numbers density ($N_p$), and plasma bulk speed ($U$). Each panel shows OMNI data (black), and simulated solar wind using real (red) and synthetic (blue) magnetogram. Panels (a-c) show the solar wind comparisons at 1 AU for CRs 1925, 1957, and 1989 respectively, within the solar cycle 23, and panels (d-i) show results for CRs 2021, 2069, 2086, 2112, 2151, and 2164 within the solar cycle 24. 

\begin{figure*}[htpb!]  
\epsscale{0.60}
\plotone{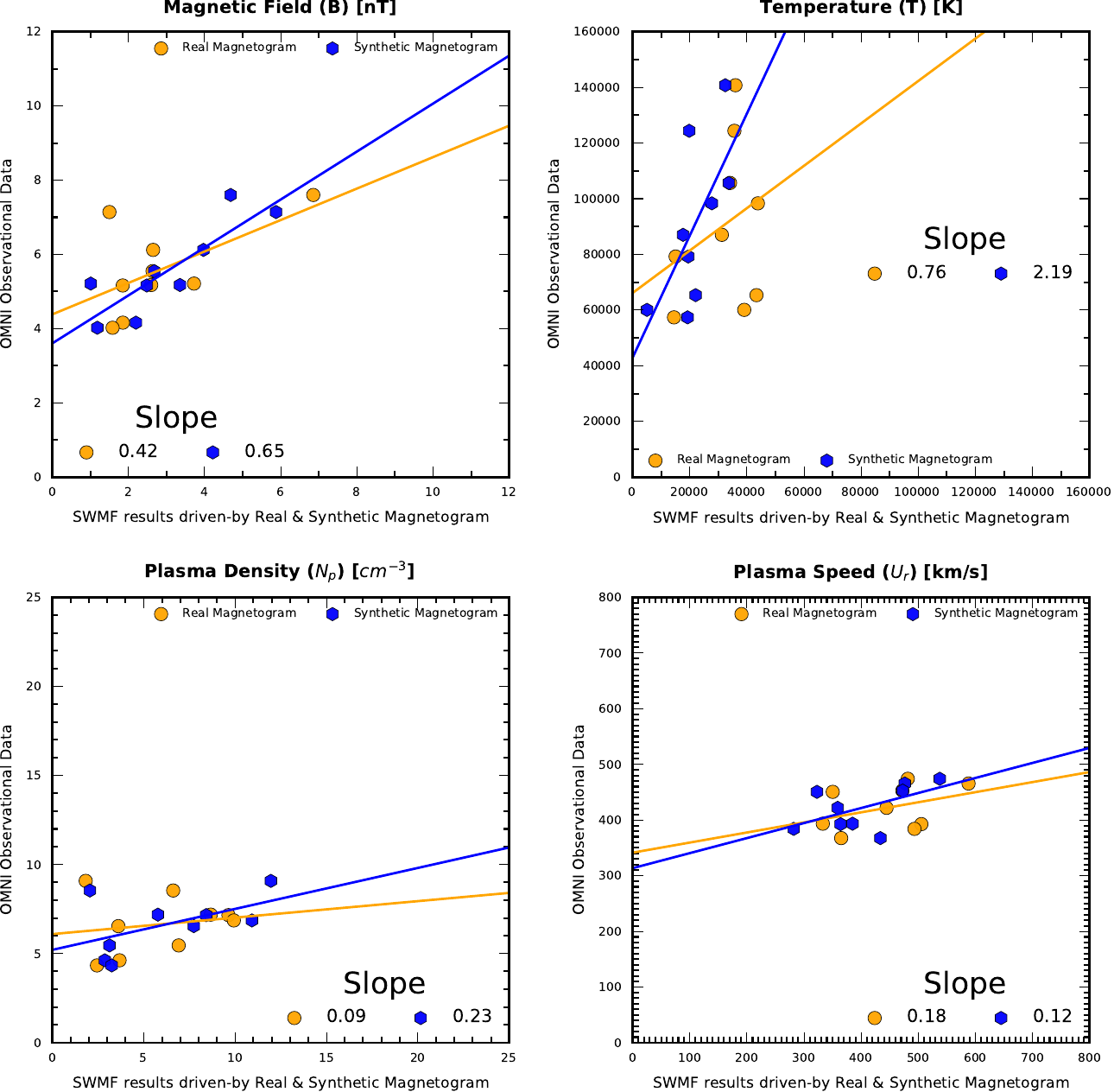}
\caption{Correlation between the mean values of the solar wind parameters obtained from SWMF and OMNI observations at 1 AU for CRs 1925, 1957, 1989, 2021, 2069, 2086, 2112, 2151, and 2164.}
\label{fig:mean of SWP}
\end{figure*}

\begin{figure*}[htpb!]  
\epsscale{0.60}
\plotone{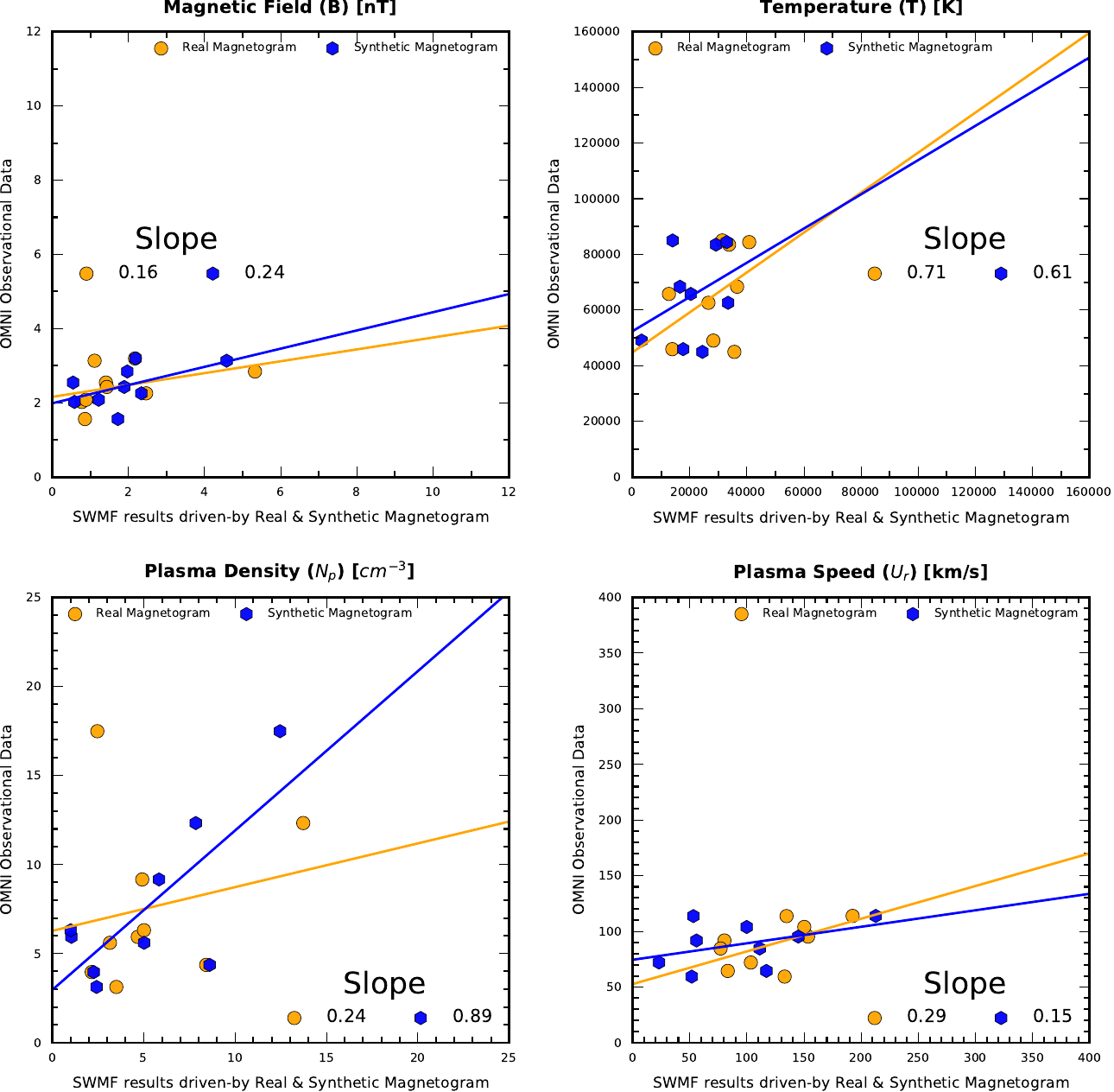}
\caption{Correlation between the standard deviations of the solar wind parameters obtained from SWMF and OMNI observations at 1 AU for CRs 1925, 1957, 1989, 2021, 2069, 2086, 2112, 2151, and 2164.
\label{fig:std of SWP}}
\end{figure*}

Despite of the fact that our goal here is not to validate the MHD model, but to focus on comparing the results from the two magnetogram inputs, one should expect a reasonable performance of SWMF. However, the 1 hour, point-by-point comparison shown in figure~\ref{fig:OMNI_linear} may undermine the model's performance, as it represents a comparison of a global model, with a rather large grid size near the Earth, with a single point of measurement in time and space. Instead, we choose to quantify the model's performance in a more statistical manner. For each data set (OMNI data, real, and synthetic pmagnetogram), each CR, and a given parameter, we create a probability distribution of the 1~AU values over the duration of the CR $\pm$ two weeks. Then, instead of comparing the time series, we compare these distributions by defining their statistical properties - their mean and standard deviation. Figure \ref{fig:DISTall} (a-i) shows the probability density distributions for each solar wind parameter obtained from SWMF results driven by real (red) and synthetic (blue) magnetogram, compared with OMNI data (black) for the selected CRs. For each distribution, we calculate the mean values and the standard deviation to measure the spread and center point of the particular case and parameter. Tables \ref{tab:Mean values} and \ref{tab:Standard deviation} summarize the mean and standard deviation values. Figure \ref{fig:mean of SWP} shows the mean values of each parameter of the distributions for SWMF results plotted against those of the OMNI observations. Figure \ref{fig:std of SWP} shows similar plots for the standard deviation.  

\begin{deluxetable*}{@{\extracolsep{6pt}}ccccccccccccc}[htpb!]
\tablecaption{Mean values of the distribution for solar wind parameters at 1 AU. \label{tab:Mean values}}
\tablewidth{0pc}
\tabletypesize{\footnotesize}
\tablehead
{
\colhead{}&
  \multicolumn{4}{c}{OMNI Observations}&
  \multicolumn{4}{c}{SWMF(Input:Real Magnetogram)}&
  \multicolumn{4}{c}{SWMF(Input:Synthetic Magnetogram)}\\
\cline{2-5} \cline{6-9} \cline{10-13} 
\colhead{}& 
\colhead{B}& \colhead{T}& \colhead{$N_p$}& \colhead{$U_r$}& 
\colhead{B}& \colhead{T}& \colhead{$N_p$}&
\colhead{$U_r$}& 
\colhead{B}& \colhead{T}& \colhead{$N_p$}& \colhead{$U_r$}\\
\colhead{CR} & 
\colhead{(nT)} & \colhead{(K)} &
\colhead{($cm^{-3}$)} & \colhead{(km/s)} & 
\colhead{(nT)} & \colhead{(K)} & \colhead{($cm^{-3}$)} & \colhead{(km/s)} & 
\colhead{(nT)} & \colhead{(K)} & \colhead{($cm^{-3}$)} & \colhead{(km/s)}
}
\startdata
\vspace{0.4em}
1925 & 5.21 & 65351 & 8.40 & 390.98 & 3.68 & 42933 & 6.64 & 503.66 & 1.01 & 21857 & 2.06 & 363.26\\
\vspace{0.4em}
1957 & 7.14 & 140764 & 5.19 & 474.05 & 1.53 & 36970 & 1.83 & 487.35 & 5.71 & 32911 & 11.52 & 543.42\\
\vspace{0.4em}
1989 & 7.60 & 124390 & 5.43 & 450.56 & 6.87 & 35203 & 9.58 & 348.72 & 4.69 & 19654 & 8.35 & 322.62\\
\vspace{0.4em}
2021 & 5.55 & 79191 & 6.24 & 391.60 & 2.62 & 15297 & 8.55 & 334.37 & 2.64 & 19190 & 5.74 & 382.17\\
\vspace{0.4em}
2069 & 4.16 & 98334 & 4.62 & 465.53 & 1.85 & 43843 & 3.76 & 586.88 & 2.19 & 29317 & 2.84 & 483.17\\
\vspace{0.4em}
2086 & 4.02 & 60093 & 5.08 & 384.09 & 1.56 & 38018 & 7.07 & 488.81 & 1.20 & 5046 & 3.20 & 281.58\\
\vspace{0.4em}
2112 & 5.16 & 105658 & 4.26 & 452.72 & 1.85 & 34156 & 2.45 & 472.22 & 2.48 & 33645 & 3.25 & 472.85\\
\vspace{0.4em}
2151 & 5.18 & 57371 & 6.86 & 367.48 & 2.60 & 14470 & 9.93 & 365.01 & 3.35 & 19231 & 10.92 & 433.67\\
\vspace{0.4em}
2164 & 6.08 & 86197 & 6.66 & 417.58 & 2.64 & 31530 & 3.59 & 445.74 & 3.97 & 17692 & 7.74 & 358.13\\
\hline
\enddata
\end{deluxetable*}

\begin{deluxetable*}{@{\extracolsep{6pt}}ccccccccccccc}[htpb!]
\tablecaption{Standard deviation of the distribution for solar wind parameters at 1 AU. \label{tab:Standard deviation}}
\tablewidth{0pc}
\tabletypesize{\footnotesize}
\tablehead
{
\colhead{}&
  \multicolumn{4}{c}{OMNI Observations}&
  \multicolumn{4}{c}{SWMF(Input:Real Magnetogram)}&
  \multicolumn{4}{c}{SWMF(Input:Synthetic Magnetogram)}\\
\cline{2-5} \cline{6-9} \cline{10-13} 
\colhead{}& 
\colhead{B}& \colhead{T}& \colhead{$N_p$}& \colhead{$U_r$}& 
\colhead{B}& \colhead{T}& \colhead{$N_p$}& \colhead{$U_r$}& 
\colhead{B}& \colhead{T}& \colhead{$N_p$}&
\colhead{$U_r$}\\
\colhead{CR} & 
\colhead{(nT)} & \colhead{(K)} &
\colhead{($cm^{-3}$)} & \colhead{(km/s)} & 
\colhead{(nT)} & \colhead{(K)} & \colhead{($cm^{-3}$)} & \colhead{(km/s)} & 
\colhead{(nT)} & \colhead{(K)} & \colhead{($cm^{-3}$)} & \colhead{(km/s)}
}
\startdata
\vspace{0.4em}
1925 & 2.55 & 44974 & 5.00 & 46.96 & 1.42 & 35569 & 4.65 & 132.89 & 0.54 & 24282 & 1.05 & 51.50\\
\vspace{0.4em}
1957 & 3.13 & 83490 & 5.54 & 113.86 & 1.11 & 33712 & 2.50 & 194.01 & 4.61 & 29384 & 12.32 & 212.32\\
\vspace{0.4em}
1989 & 2.84 & 84996 & 5.11 & 91.78 & 5.29 & 31132 & 13.58 & 80.16 & 1.95 & 13847 & 7.77 & 55.22\\
\vspace{0.4em}
2021 & 2.42 & 65720 & 3.82 & 76.63 & 1.42 & 13196 & 4.91 & 78.36 & 1.86 & 20042 & 5.72 & 109.66\\
\vspace{0.4em}
2069 & 1.56 & 62607 & 3.13 & 95.30 & 0.85 & 26743 & 3.62 & 154.80 & 1.71 & 36380 & 2.41 & 150.65\\
\vspace{0.4em}
2086 & 2.02 & 49045 & 3.48 & 72.05 & 0.77 & 28219 & 5.01 & 105.04 & 0.60 & 3087 & 1.10 & 23.40\\
\vspace{0.4em}
2112 & 2.08 & 84401 & 3.13 & 103.93 & 0.88 & 40781 & 2.15 & 150.14 & 1.22 & 32947 & 2.27 & 99.74\\
\vspace{0.4em}
2151 & 2.26 & 45943 & 4.37 & 64.44 & 2.46 & 13877 & 8.42 & 83.23 & 2.34 & 17685 & 8.61 & 117.05\\
\vspace{0.4em}
2164 & 3.21 & 67643 & 5.62 & 111.47 & 2.17 & 36611 & 3.17 & 135.80 & 2.19 & 16573 & 5.02 &
52.90\\
\hline
\enddata
\end{deluxetable*}

\section{Discussion} 
\label{sec:discussion}

This paper introduces surface field maps (Synthetic magnetogram) produced by the 3D kinematic dynamo (Kd3) code as a new input for solar wind MHD simulations. To validate the new input we first optimize the SWMF for the real magnetogram input. We have shown that this input could potentially reproduce the solar wind parameters at 1 AU and the LOS EUV images with a reasonable agreement with predictions made by real magnetograms for some CRs. 

When looking at the comparisons obtained for LOS EUV images (see figure \ref{fig:EUV_all}), we find that the synthetic magnetogram could reproduce the EUV images in reasonable agreement with observations. However, it could not provide much detailed information on the lower corona as the real magnetogram can. In general, EUV bright regions are produced by the magnetogram's active regions, and the dark coronal holes are produced by the open magnetic field regions, typically dictated by the lower order magnetogram component. When considering EUV comparison for CR 1925 (figure \ref{fig:EUV_all}a.) centered on {\it{1997.07.29}}, it is quite clear that the real magnetogram provides a better result than the synthetic magnetogram. However, the synthetic magnetogram result captures the bright feature clearly on the limb, as seen in the observations. On the other hand, for the CR 2086 comparison (figure \ref{fig:EUV_all}f.) centered on {\it{2009.08.06}}, the synthetic magnetogram reproduces additional bright features, which are not in SOHO/EIT observations. However, the real magnetogram result perfectly captures the observed images while reproducing the coronal hole morphology and the bright feature observed from 1 AU. The overestimation here is because of the uncertainty of the Kd3 code in reproducing the bright features of the sun at a period with very few active regions. So, when imposing velocity perturbations, the code amplifies the tiny bright spots of the sun as large plumes.

Moreover, considering the EUV image for CR 2069 (figure \ref{fig:EUV_all}e.) centered on {\it{2008.04.29}}, we see that the synthetic magnetogram result was unable to capture the coronal hole morphology quite well, as shown in the real magnetogram simulated images. However, it captures the bright feature on the limb as observed in SOHO/EIT images. The above mentioned three rotations are for a period of solar-minimum, occurred in solar cycle 23 (CR 1925), and 24 (CR 2069, 2086). All the other rotations (occurred during a solar-maximum) show a reasonable agreement with real and synthetic magnetograms in terms of the active regions and coronal hole morphology. Looking at the comparison for CR 1957 (figure \ref{fig:EUV_all}b.) centered on {\it{1999.12.18}}, both SWMF synthesized EUV images show an offset in the location of the bright features of the sun. Other than that, all other rotations (figure \ref{fig:EUV_all}c, d, g, h, and i) show a reasonable agreement, with the synthetic magnetogram images being a blurry version of the real magnetogram images. This is expected as the synthetic magnetograms do not provide sharp, clear active regions. We plan to dedicate a future study to investigate how the sharpness of the active regions in the synthetic magnetograms can be improved, perhaps with an increased model resolution or a post-processing image sharpening of the output magnetogram.

\begin{figure*}[htpb]  
\plotone{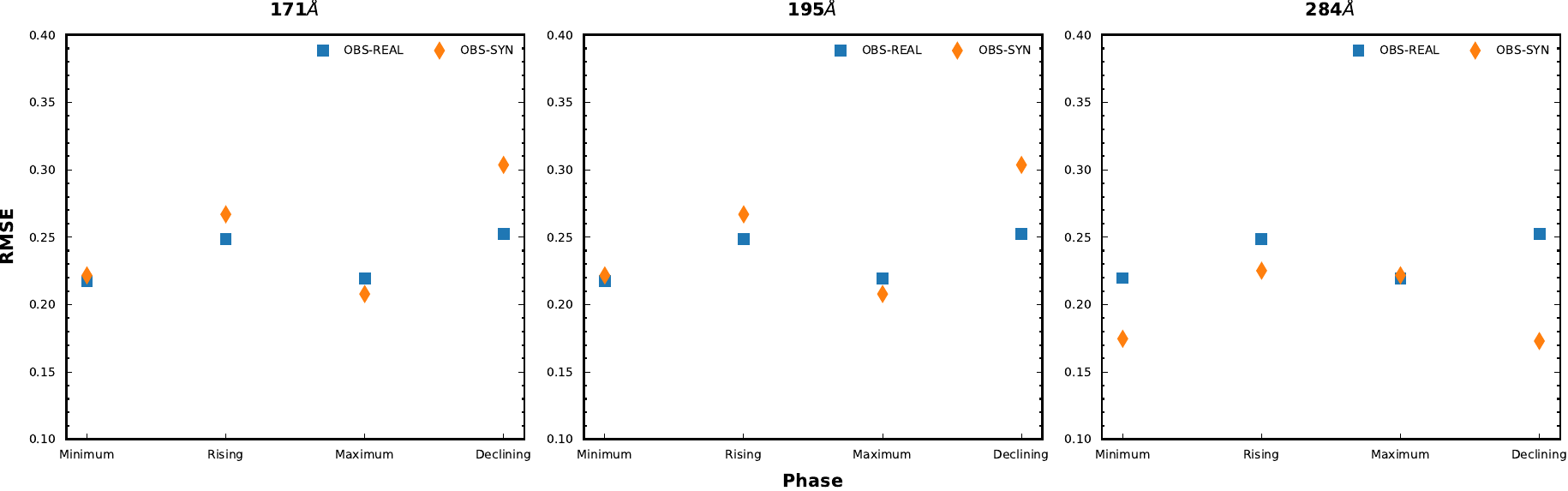}
\caption{Average RMSE of EUVI for (left) 171\;\AA \; (middle) 195\;\AA \;and (right) 284\;\AA.
\label{fig:RMSE}}
\end{figure*}

To observe the comparison statistically, we performed calculations for the Root Mean Square Error (RMSE), which compares the quantitative behavior of the EUVI between the model synthesized and the observed images. We refer the reader to table \ref{tab:RMSE} for the values of RMSE. From the table \ref{tab:RMSE} we see that the RMSE for the model synthesized images driven using the real and the synthetic magnetograms are in excellent agreement. The RMSE between the observed and the model synthesized images using a synthetic magnetogram is slightly higher than the real magnetogram. However, for all cases, the RMSE values are in order of less than $\sim$ 0.3, implying the model captures the coronal hole morphology and the bright regions of the corona in good agreement with the SOHO/EIT observations for both real and synthetic magnetogram inputs. Further, to observe how the EUVI RMSE comparison stands for each phase of the solar cycle, we categorized the nine rotations we used for the simulations under each phase (see table \ref{tab:Phase}). Here we performed an average RMSE analyses for EUV bands 171 \AA, 195 \AA, and 284\;\AA (see figure \ref{fig:RMSE}).


\begin{deluxetable}{cc}[!htpb]
\tablecaption{CRs and the phase of the solar cycle. \label{tab:Phase}}
\tablewidth{0pt}
\tablehead{\colhead{Phase of the Solar cycle} & \colhead{CRs} 
}
\startdata
Minimum & 1925,\;2069,\;2086 \\
\hline
Rising & 1957,\;2112 \\
\hline
Maximum & 1989,\;2151  \\
\hline
Declining & 2021,\;2164 \\
\hline
\enddata
\end{deluxetable}

From the figure \ref{fig:RMSE}, we see that for all bands, the RMSE is higher for the rising phase of the solar cycle, overall rising and the maximum phase showing consistent results for both real and synthetic magnetogram cases. Interestingly, the RMSE values obtained for the synthetic magnetogram in the 284\;\AA\ band is lower than for the real magnetogram comparison. From the quantitative point of view, for all phases, we see that the synthetic magnetogram does a better job synthesizing the EUVI than the real magnetogram in the 284\;\AA\ band. The 195\;\AA\ result implies that the real and synthetic magnetogram synthesizes the EUVI consistently with the observations from SOHO/EIT. From the observations point of view, we see that the synthetic magnetogram synthesized images are worse than the observed and the real magnetogram synthesized images. However, the quantitative comparison shows that the model can reproduce the EUVI in excellent agreement for both real and synthetic magnetogram inputs.

\begin{deluxetable}{@{\extracolsep{6pt}}ccccccccccccc}[htpb]
\centering
\tablecaption{Root Mean Square Error (RMSE) for EUVI. \label{tab:RMSE}}
\tablewidth{0pc}
\tabletypesize{\footnotesize}
\tablehead
{
\colhead{}&
  \multicolumn{4}{c}{RMSE}\\
\cline{2-5} 
\colhead{CR}& 
\colhead{Wavelength (\AA)}& \colhead{Obs-Real}& \colhead{Obs-Syn}& \colhead{Real-Syn}
}
\startdata
& 171 & 0.1913 & 0.2388 & 0.1309\\
1925 & 195 & 0.1579 & 0.1899 & 0.1235 \\
& 284 & 0.1581 & 0.1667 & 0.0917 \\
\hline
& 171 & 0.2760 & 0.2881 & 0.2184\\
1957 & 195 & 0.2280 & 0.2436 & 0.1628 \\
& 284 & 0.2414 & 0.2570 & 0.1169 \\
\hline
& 171 & 0.1594 & 0.1748 & 0.1303\\
1989 & 195 & 0.2068 & 0.2429 & 0.0957 \\
& 284 & 0.2337 & 0.2418 & 0.1128 \\
\hline 
& 171 & 0.2435 & 0.2669 & 0.1230\\
2021 & 195 & 0.1484 & 0.1902 & 0.1092\\
& 284 & 0.1615 & 0.1747 & 0.0902\\
\hline
& 171 & 0.2143 & 0.1831 & 0.1377\\
2069 & 195 & 0.1349 & 0.1649 & 0.1010\\
& 284 & 0.1463 & 0.1588 & 0.0802\\
\hline
& 171 & 0.2471 & 0.2428 & 0.2337\\
2086 & 195 & 0.1564 & 0.1995 & 0.1553 \\
& 284 & 0.1599 & 0.1984 & 0.1296 \\
\hline
& 171 & 0.2215 & 0.2457 & 0.1628\\
2112 & 195 & 0.1595 & 0.1996 & 0.1321 \\
& 284 & 0.1651 & 0.1932 & 0.1318 \\
\hline
& 171 & 0.2793 & 0.2408 & 0.2023\\
2151 & 195 & 0.1711 & 0.1782 & 0.1165 \\
& 284 & 0.1980 & 0.2015 & 0.1339\\
\hline
& 171 & 0.2607 & 0.3403 & 0.1699\\
2164 & 195 & 0.1785 & 0.1897 & 0.1029\\
& 284 & 0.1691 & 0.1710 & 0.1216\\
\hline
\enddata
\end{deluxetable}

\begin{figure*}[htpb]  
\epsscale{1.17} 
\plotone{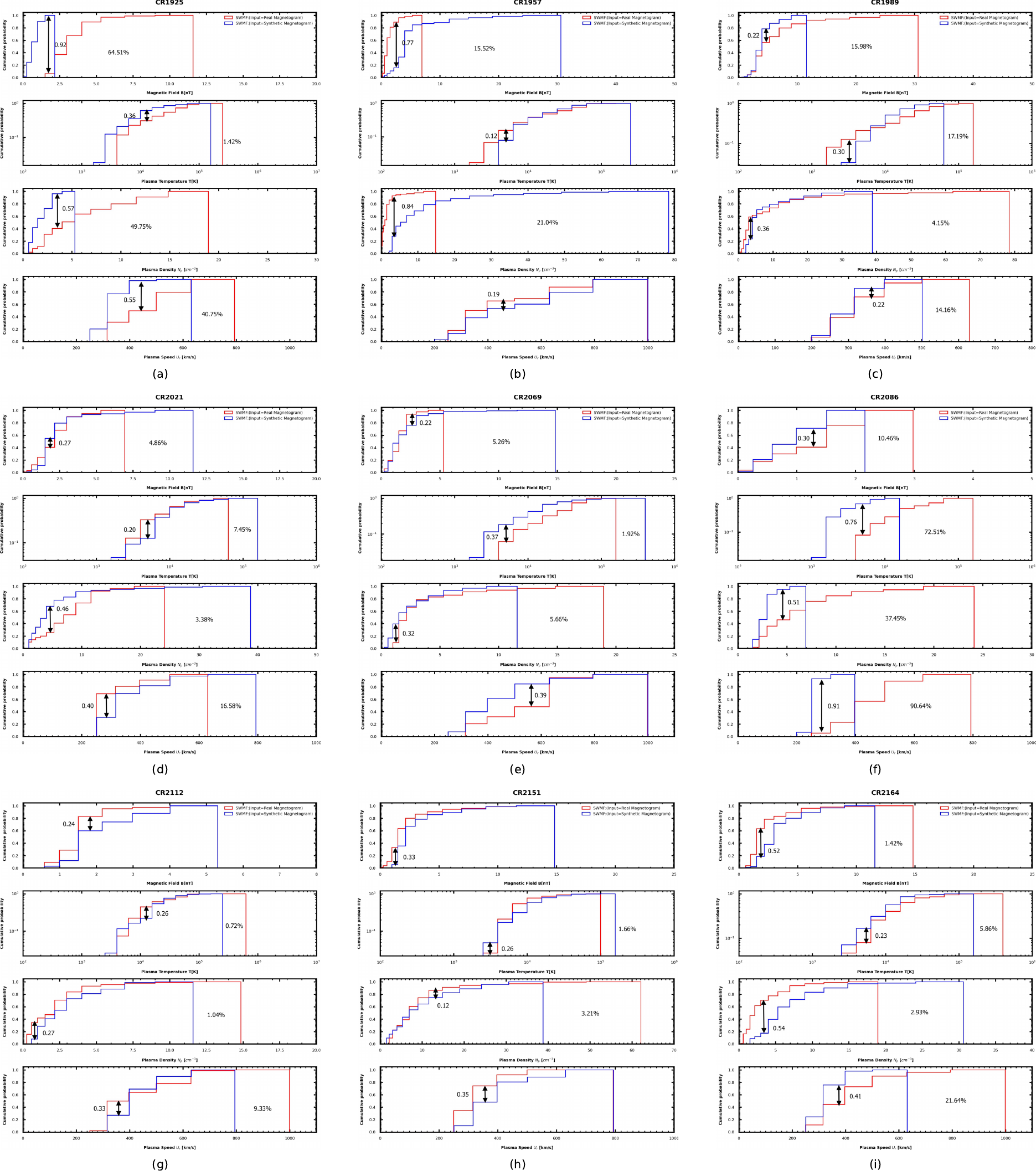}
\caption{Comparison of the cumulative probability distributions for solar wind at 1 AU. SWMF results for solar wind parameters driven by real magnetogram data (red) and from synthetic magnetogram data (blue) for CRs, (a) 1925, (b) 1957, (c) 1989, (d) 2021, (e) 2069, (f) 2086, (g) 2112, (h) 2151, and (i) 2164.
\label{fig:ks}}          
\end{figure*}

The SWMF results show that both real and synthetic magnetogram input statistically reproduces the observed solar wind parameters at 1 AU reasonably well for most CRs (see the tables \ref{tab:Mean values} \& \ref{tab:Standard deviation} for the mean and standard deviation of the distributions). The SWMF results for CR 1925 and 2069 (solar minimum, see figure \ref{fig:OMNI_linear}a, and e) driven by real magnetogram produce solar wind that is faster than observed in OMNI observations, where this result is confirmed from the distributions plots as well (see figure \ref{fig:DISTall}a, and \ref{fig:DISTall}e). Interestingly, the synthetic magnetograms for these CRs produce fast and slow wind, with a better agreement with observations comparing to the real magnetogram. These solar minimum results indicate that the synthetic magnetogram could potentially reproduce the large-scale solar wind structure at 1~AU (which is affected less by the active regions) quite well.

For the magnetic field comparison, we mostly find that the SWMF results underestimate the observations, which is consistent with \cite[e.g.,][]{Cohen2008,Sachdeva_2019}. However, looking at the distributions for the magnetic field, the SWMF results obtained from the real and synthetic magnetogram show similar distributions. Real and synthetic magnetogram simulation results also show an overall underestimation of the observed solar wind temperature. This is likely due to the fact that the modeled solar wind temperature represents the single-fluid MHD temperature, which may be different than the observed OMNI proton temperature. The plasma density comparison shows that for all the CRs, the real and synthetic magnetogram-driven SWMF results show pretty good agreement. Figure \ref{fig:DISTall} shows the summarized distributions of the solar wind parameters for the CRs representing solar cycle 23 and 24. Looking at the distributions, we find that the simulation results driven by the synthetic and real magnetogram match reasonably well except for a few rotations. For CRs 1925 (figure \ref{fig:DISTall}a.) and 2086 (figure \ref{fig:DISTall}f.), we see that the synthetic magnetogram underestimates the solar wind parameters at 1 AU compared with the real magnetogram input, which proves that the synthetic magnetogram performs well when there is a fair number of active regions on the solar surface (solar-maximum conditions). These results are consistent with the EUV image comparisons and the linear plots shown in Figures \ref{fig:EUV_all}, and \ref{fig:OMNI_linear}. For more quantitative description of the distributions, please refer to table \ref{tab:Mean values} \& \ref{tab:Standard deviation}.  

\begin{deluxetable}{@{\extracolsep{6pt}}ccccccccccccc}[!ht]
\centering
\tablecaption{Kolmogorov–Smirnov test statistics between the cumulative distributions for Real and Synthetic magnetogram results. \label{tab:KS}}
\tablewidth{0pc}
\tablehead
{
\colhead{}&
  \multicolumn{4}{c}{Test statistics D}\\
\cline{2-5} 
\colhead{}& 
\colhead{B}& \colhead{T}& \colhead{$N_p$}& \colhead{$U_r$}\\
\colhead{CR} & 
\colhead{(nT)} & \colhead{(K)} & \colhead{($cm^{-3}$)} & \colhead{(km/s)}
}
\startdata
\vspace{0.4em}
1925 & 0.92 & 0.36 & 0.57 & 0.55\\
\vspace{0.4em}
1957 & 0.77 & 0.12 & 0.84 & 0.19 \\
\vspace{0.4em}
1989 & 0.22 & 0.30 & 0.36 & 0.22 \\
\vspace{0.4em}
2021 & 0.27 & 0.20 & 0.46 & 0.40 \\
\vspace{0.4em}
2069 & 0.22 & 0.37 & 0.32 & 0.39 \\
\vspace{0.4em}
2086 & 0.30 & 0.76 & 0.51 & 0.91 \\
\vspace{0.4em}
2112 & 0.24 & 0.26 & 0.27 & 0.33 \\
\vspace{0.4em}
2151 & 0.33 & 0.26 & 0.12 & 0.35 \\
\vspace{0.4em}
2164 & 0.52 & 0.23 & 0.54 & 0.41 \\
\hline
\enddata
\end{deluxetable}

\begin{deluxetable}{@{\extracolsep{6pt}}ccccccccccccc}[!ht]
\centering
\tablecaption{Wasserstein distance between the distributions for Real and Synthetic magnetogram results. \label{tab:WSD}}
\tablewidth{0pc}
\tablehead
{
\colhead{}&
  \multicolumn{4}{c}{Solar wind parameters}\\
\cline{2-5} 
\colhead{}& 
\colhead{B}& \colhead{T}& \colhead{$N_p$}& \colhead{$U_r$}\\
\colhead{CR} & 
\colhead{(nT)} & \colhead{(K)} & \colhead{($cm^{-3}$)} & \colhead{(km/s)}
}
\startdata
\vspace{0.4em}
1925 & 2.68 & 21076.16 & 4.58 & 140.40\\
\vspace{0.4em}
1957 & 4.18 & 5069.85 & 9.69 & 60.36\\
\vspace{0.4em}
1989 & 2.20 & 16617.00 & 3.01 & 30.79\\
\vspace{0.4em}
2021 & 0.59 & 4408.33 & 3.40 & 49.88\\
\vspace{0.4em}
2069 & 0.54 & 20973.92 & 0.98 & 107.61\\
\vspace{0.4em}
2086 & 0.38 & 32972.21 & 3.87 & 207.23\\
\vspace{0.4em}
2112 & 0.62 & 8579.07 & 0.87 & 48.74\\
\vspace{0.4em}
2151 & 0.85 & 4879.25 & 1.98 & 68.66\\
\vspace{0.4em}
2164 & 1.38 & 13680.20 & 4.10 & 88.11\\
\hline
\enddata
\end{deluxetable}

To further investigate the statistical performance of the model, we perform two additional statistical tests, the Kolmogorov Smirnov (KS) test and the Wasserstein distance calculations (Earth movers' distance). The KS test measures the degree of separation between the cumulative distribution of the two samples. In this work, we selected the null hypothesis $H_0$ by stating that the real and synthetic magnetogram simulated distributions are identical. We calculated test statistics {\it D} (see table \ref{tab:KS}), and we found that the obtained {\it p-values} for each solar wind parameter is $\approx$ 0, which rejects the $H_0$ stating the alternative hypothesis $H_a$ is true. In other words, the KS test states that the distributions are not identical or not from a similar population. We calculated the test statistics using the {\it scipy.stats.ks2\_samp}\footnote{{\tt https://docs.scipy.org/doc/scipy/\\reference/generated/scipy.stats.ks_2samp.html}} Python module. Even though the results from the KS test did not confirm promising similarities of the two distributions, we identify a positive trend by looking at the cumulative distribution plots, for where we found that the calculated values for test statistics {\it D} are small\;(see table \ref{tab:KS}).  

Figures \ref{fig:ks} shows the cumulative distributions for all CRs representing the solar wind parameters, including the values of D statistics. The D statistics provide the maximum vertical difference between the real and synthetic magnetogram-driven cumulative distributions. The statistics for each CR summarizes in Table~\ref{tab:KS}. In figure \ref{fig:ks}, we have interpreted the percentages by how much the distributions are off to the maximum of the other. The worst-case result is for the temperature distribution in CR 2086 (figure \ref{fig:ks}f.), and the observed value is 72.51 \%. Further investigating the cumulative distributions, we identify that the results we stated earlier for CR 1925 (figure \ref{fig:ks}a.) and 2086 (figure \ref{fig:ks}f.) are consistent with linear and probability density distribution plots. Overall, the cumulative distributions show that both inputs reproduce plasma speed and density at 1 AU in a good agreement. Cumulative distributions for CRs 2112 (figure \ref{fig:ks}g.), and 2151 (figure \ref{fig:ks}h.) show exceptional agreement to each other, thus supporting our previous finding that synthetic magnetograms reproduce the solar wind at 1 AU better during solar-maximum.

\begin{figure*}[htpb]  
\epsscale{1.17} 
\plotone{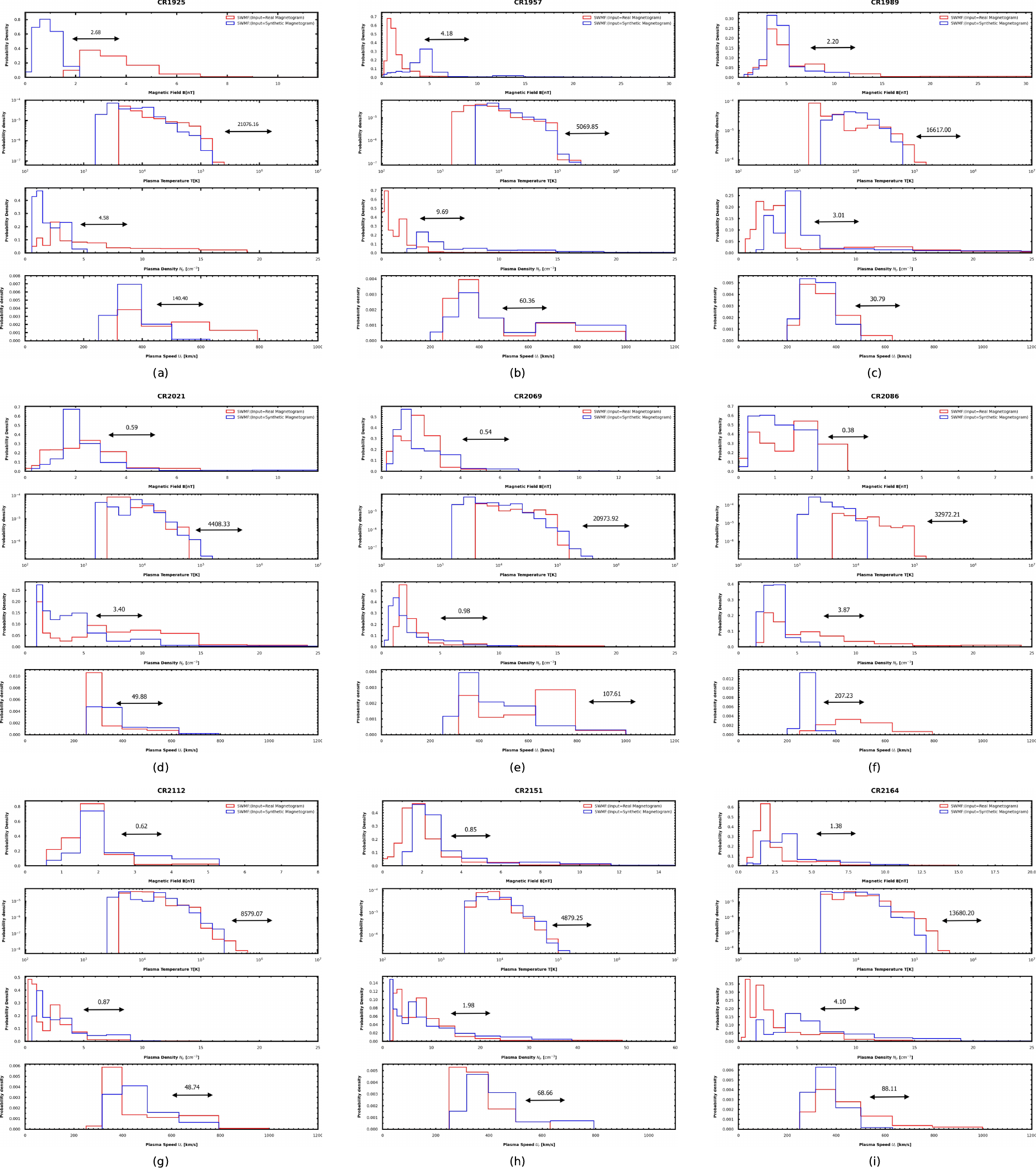}
\caption{Comparison of the distributions for solar wind at 1 AU. SWMF results for solar wind parameters driven by real magnetogram data (red) and from synthetic magnetogram data (blue) for CRs, (a) 1925, (b) 1957, (c) 1989, (d) 2021, (e) 2069, (f) 2086, (g) 2112, (h) 2151, and (i) 2164.
\label{fig:WSD}}         
\end{figure*}

To test the two probability density distributions in the horizontal direction, we perform the Wasserstein distance (Earth movers’ distance) calculations (see table \ref{tab:WSD}). This test deduces the minimum work where one distribution requires to change into the other [\citep{2021ApJ...922...15C} \& \citep{2021arXiv210602051L}]. To calculate the distance, we use {\it scipy.stats.wasserstein\_distance}\footnote{{\tt https://docs.scipy.org/doc/scipy/\\reference/generated/scipy.stats.wasserstein_distance.html}} python module. The comparison was for the simulation results obtained from the real and synthetic magnetograms. The values for each case display on the figure \ref{fig:WSD}, and in table \ref{tab:WSD}. From, table \ref{tab:WSD} we see that the Earth movers distance values for the solar wind speed for CR 1925, 2069, and 2086 which occurred in a solar minimum condition are high compared to the other rotations occurred in a solar maximum.

These calculations confirmed the underestimation of the synthetic magnetogram input in reproducing solar wind at 1 AU for CRs 1925 (figure \ref{fig:WSD}a.) and 2086 (figure \ref{fig:WSD}f.). Overall, we see a reasonable agreement in both distributions for the other rotations. The calculated Wasserstein distances are small compared to the scale of each solar wind parameter, which means the real and synthetic magnetogram distributions show good agreement when quantifying in the horizontal direction.

Observing all the statistical results, we see that the simulated solar wind parameters obtained for the CRs representing solar cycle 24 show better agreement with observations than those in solar cycle 23. We obtained the mean values and the standard deviations for each solar wind parameter from the distributions (see table \ref{tab:Mean values} and \ref{tab:Standard deviation}). Figure \ref{fig:mean of SWP} and \ref{fig:std of SWP} shows the linear relations for the mean and standard deviations for each solar wind parameter obtained for CRs 1925, 1957, 1989, 2021, 2069, 2086, 2112, 2151, and 2164. We do not see a strong correlation between the model simulated solar wind parameters with OMNI observations from the plots. However, by looking at the relatively close slopes of both figures \ref{fig:mean of SWP} and \ref{fig:std of SWP}, we see that there is a good linear relationship between the real and synthetic magnetogram results.

One of the primary concerns in this comparison is having to use magnetograms from different observatories due to the lack of observed data for the periods of interest. The magnetograms from the same observatory might help keep the comparison consistent for all cases. 

Overall, our results show that when the synthetic magnetograms preform well in terms of reproducing the coronal images when there are more active regions on the disk. For solar minimum conditions, the synthetic magnetograms seem to lack the details needed to reproduce those images. However, for 1~AU solar wind conditions, the synthetic magnetograms show the potential to statistically perform almost as well as the real magnetograms in predicting the ambient solar wind. Therefore, it show the potential to provide predictions for the ambient solar wind prior to the actual state of the Sun and the Heliosphere.

\section{Conclusion} \label{sec:Conclusion}
We study whether synthetic magnetograms can be use to predict the ambient solar wind at 1~AU by comparing it's performance against real magnetogram data. The comprehensive study and analysis conclude that synthetic magnetograms could initialize solar wind models for future space weather predictions as an alternative to the observed magnetograms in reproducing realistic solar wind conditions at 1 AU. Overall looking at the results, we conclude that the synthetic magnetogram performs better in reproducing 1 AU results when a certain number of active regions are present on the solar surface or in a solar maximum. We plan to dedicate future study to investigate how synthetic magnetograms can be improved to provide better predictions for the solar wind at 1~AU.

\begin{center} 
ACKNOWLEDGMENTS
\end{center}
We thank an unknown referee for their useful comments. This work is supported by NASA HSR grant 80NSSC20K1354. Simulation results were obtained using the (open source) Space Weather Modeling Framework, developed by the Center for Space Environment Modeling, at the University of Michigan with funding support from NASA ESS, NASA ESTO-CT, NSF KDI, and DoD MURI. The simulations were performed on the Massachusetts Green High-Performance Computing Center (MGHPCC) cluster supercomputer. We thank Dr.Soumitra Hazra for his suggestions which improved this paper.



\end{document}